\documentclass[12pt]{amsart}
\usepackage{amsfonts}
\usepackage{amssymb,mathrsfs}
\usepackage{amsmath}
\usepackage{enumerate}
\usepackage{latexsym}
\usepackage{graphicx}
\usepackage{bm}
\usepackage{subfigure}
\usepackage{float}
\usepackage{amsmath}
\usepackage{amsthm}
\usepackage{verbatim}
\usepackage{vmargin}
\usepackage{amstext}
\usepackage{array}
\usepackage{booktabs}
\usepackage{indentfirst}
\usepackage[compress]{cite}
\usepackage{url}
\usepackage{geometry}
\numberwithin{equation}{section}

\newtheorem{thm}{Theorem}[section]
\newtheorem{cor}[thm]{Corollary}

\newtheorem{rem}[thm]{Remark}

\theoremstyle{definition}

\begin{document}
\setlength{\oddsidemargin}{80pt}
\setlength{\evensidemargin}{80pt}
\setlength{\topmargin}{80pt}
\author{Jun~Xian, Xiaoda~Xu}

\address{J.~Xian\\Department of Mathematics and Guangdong Province Key Laboratory of Computational Science
 \\Sun
Yat-sen University
 \\
510275 Guangzhou\\
China.} \email{xianjun@mail.sysu.edu.cn}

\address{X.~Xu\\Department of Mathematics
 \\Sun
Yat-sen University
 \\
510275 Guangzhou\\
China.} \email{xuxd26@mail2.sysu.edu.cn}

\title[Expected $L_2$-discrepancy for stratified sampling]{Improved expected $L_2$-discrepancy formulas on jittered sampling}
 
\date{\today}

\pdfoutput=1

\begin{abstract}
We study the expected $ L_2-$discrepancy under two classes of partitions, explicit and exact formulas are derived respectively. These results attain better expected $L_2-$discrepancy formulas than jittered sampling.
\end{abstract}

\maketitle
\section{Introduction}\label{intro}
Among various techniques for measuring the irregularity of point distribution, discrepancy methods have proven to be one of the most efficient approaches. In various notions of discrepancy, $L_2-$discrepancy is widely studied, see applications in areas of low discrepancy point sets \cite{Dick2005,Dick2006,Dick2014,Dick2020}, and the best known asymptotic upper bounds of $L_2$-discrepancy for these point sets are of the form $$O(\frac{(\ln N)^{\frac{d-1}{2}}}{N}).$$

In order to facilitate the comparison of $L_2$-discrepancy for different sampling point sets, it is necessary to derive explicit formulas for $L_2$-discrepancy of different sampling sets. Thanks to Warnock’s formula, explicit $L_2$-discrepancy formulas for deterministic point sets can be derived, see \cite{Dick2014}. The interesting thing is to calculate the expected $L_2-$discrepancy formula for random sampling, this problem comes from the integration approximation, which is, for $f\in\mathcal{H}^{\mathbf{1}}(K)$ and random sampling set $P_\eta= \{x_i\}_{i=1}^{N}$, it is proved in \cite{Dick2014}

\begin{equation}\label{eq1d1}
\mathbb{E}[\sup_{f\in \mathcal{H}^{\mathbf{1}}(K),\|f\|_{\mathcal{H}^{\mathbf{1}}(K)}\leq 1 }\Big|\frac{1}{N}\sum_{n=1}^{N}f(x_i)-\int_{[0,1]^{d}}f(x)dx\Big|^2]\leq \mathbb{E}L_{2}^2(D_{N},P_{\eta}),\end{equation}
where $\mathcal{H}^{\mathbf{1}}(K)$ is Sobolev space.
It means that a smaller expected $L_2-$discrepancy implies a better expected uniform integration approximation in a class of functional space. Also, this means some practical approximation problems that can be converted into \eqref{eq1d1} are solvable.

The main purpose of this paper is to derive explicit $L_2-$discrepancy formulas for two classes of partitions in $d$ dimensions with sampling number $N=m^d$, and attain better results than jittered sampling under the same sampling conditions. 

The explicit formula for $L_2-$discrepancy of jittered sampling set is a known result, which is recently derived in \cite{kirk2022expected}. The significant application of improving this formula is to obtain a better and explicit upper bound of approximation error in formula \eqref{eq1d1}. Our models are motivated by \cite{KP2}, which are for $2-$dimensions with sampling sets $N=2$, and the corresponding construction models for $d-$dimensions are motivated by \cite{KP}, see model 2.4 in Section 2.

The rest of this paper is organized as follows. Section 2 presents preliminaries. Section \ref{sec3} presents our main result, which provides explicit expected $L_2$-discrepancy for two certain classes of partitions. Section \ref{Prf} includes the proofs of the main results. Finally, in section \ref{conclu} we conclude the paper with a short summary.

\section{Preliminaries on random sampling}\label{prelim}

Before introducing the main result, we list the preliminaries used in this paper.

\subsection{$L_{2}-$discrepancy} $L_{2}-$discrepancy of a sampling set $P_{N, d}=\{t_{1}, t_{2}, \ldots , t_{N}\}$ is defined by

\begin{equation}\label{lpdefn}
L_{2}(D_{N},P_{N, d})=\Big(\int_{[0,1]^{d}}|\lambda([0,z))- \frac{1}{N}\sum_{i=1}^{N}\mathbf{1}_{[0,z)}(t_{i})|^{2}dz\Big)^{1/2},
\end{equation}
where $\lambda$ denotes the Lebesgue measure, $\mathbf{1}_{A}$ denotes the characteristic function on set $A$.

In the definition of $L_2-$discrepancy, if we introduce the counting measure $\#$, \eqref{lpdefn} can also be expressed as

\begin{equation}
    L_{2}(D_{N},P_{N, d})=\Big(\int_{[0,1]^{d}}|\lambda([0,z))-
\frac{1}{N}\#\big(P_{N, d}\cap[0,z)\big)|^{2}dz\Big)^{1/2},
\end{equation}
where $\#\big(P_{N, d}\cap[0,z)\big)$ denotes the number of points falling into the set $[0,z).$

To simplify the expression of $L_{2}-$discrepancy, we employ the discrepancy function $\Delta(P_{N,d},z)$ via:

\begin{equation}\label{disfun1}
    \Delta(P_{N,d},z)=\lambda([0,z))-
\frac{1}{N}\#\big(P_{N, d}\cap[0,z)\big).
\end{equation} 

\subsection{Simple random sampling}
\begin{figure*}[h]
\centering
\subfigure[Simple random sampling in two dimensions]{
\begin{minipage}{7cm}
\centering
\includegraphics[width=0.6\textwidth]{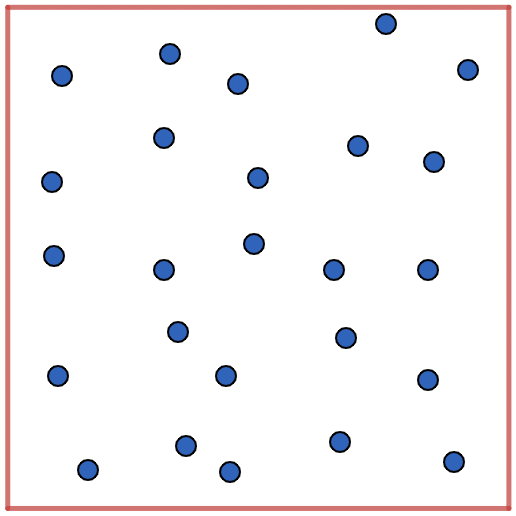}
\end{minipage}
}
\subfigure[Simple random sampling in three dimensions]{
\begin{minipage}{7cm}
\centering
\includegraphics[width=0.7\textwidth]{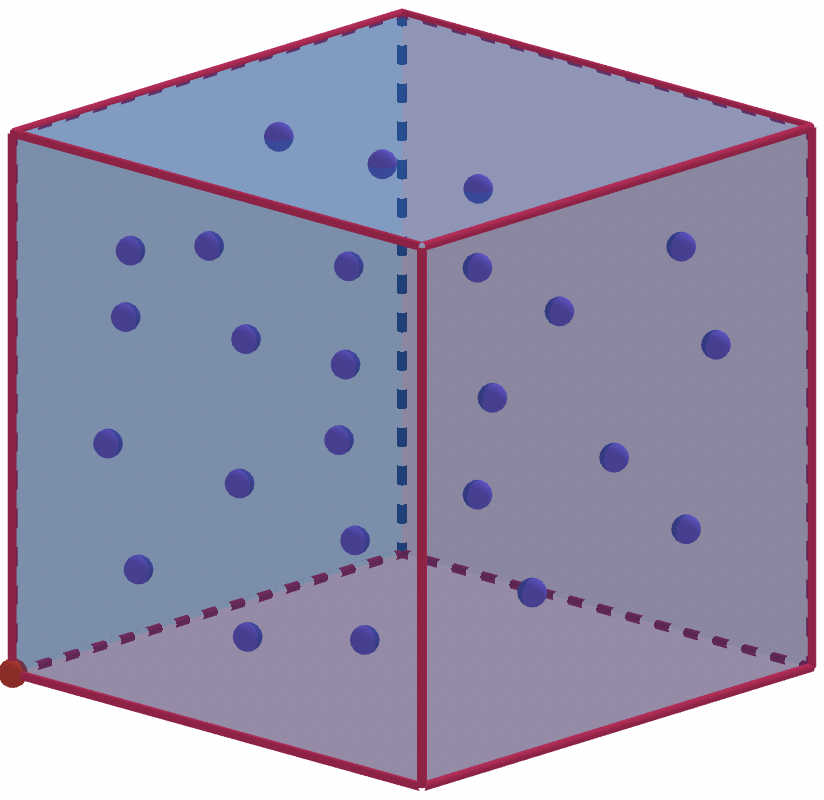}
\end{minipage}
}
\caption{\label{srs0} Simple random sampling.}
\end{figure*}

In a sense, simple random sampling is Monte Carlo sampling. Uniform distributed point set is selected in $[0,1]^{d}$, see Figure \ref{srs0}.

\subsection{Jittered sampling}
\begin{figure*}[h]
\centering
\subfigure[jittered sampling in two dimensions]{
\begin{minipage}{7cm}
\centering
\includegraphics[width=0.6\textwidth]{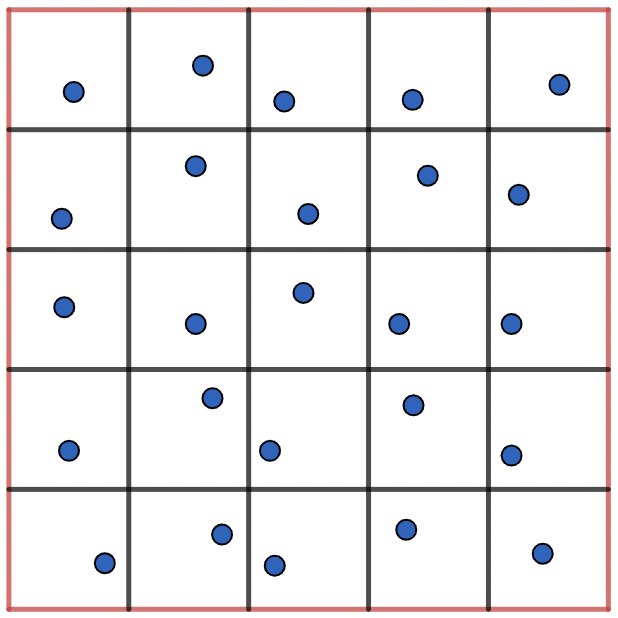}
\end{minipage}
}
\subfigure[jittered sampling in three dimensions]{
\begin{minipage}{7cm}
\centering
\includegraphics[width=0.7\textwidth]{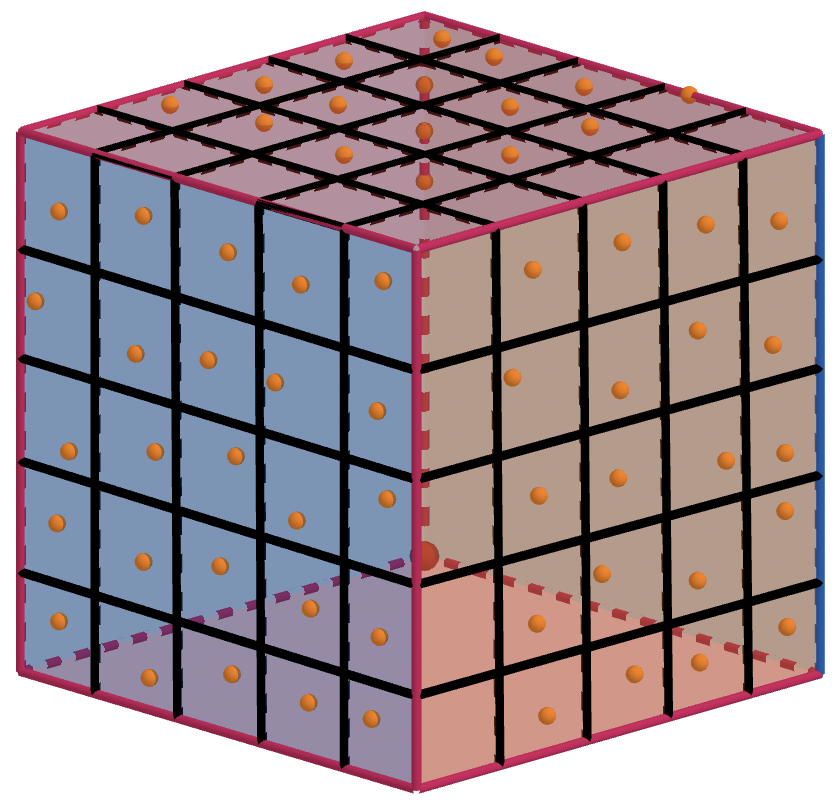}
\end{minipage}
}
\caption{\label{jss0} jittered sampling formed by isometric grid partition.}
\end{figure*}

Jittered sampling is a type of grid-based equivolume partition. $[0,1]^{d}$ is divided into $m^d$ axis parallel boxes $Q_{i},1\leq i\leq N,$ each with sides $\frac{1}{m},$ see illustration of Figure \ref{jss0}.

\subsection{Partition model in \cite{KP}}
    For a grid-based equivolume partition in two dimensions, the two squares in the upper right corner are merged to form a rectangle $$I=[a_1,a_1+2b]\times [a_2,a_2+b],$$where $a_1,a_2,b$ are three positive constants. The diagonal of $I$ is the partition line, which constitutes a special partition mode, and set 

$$\Omega_{\backslash}=(\Omega_{1,\backslash},\Omega_{2,\backslash},Q_3,\ldots,Q_{N}),$$where $\Omega_{2,\backslash}=I\setminus\Omega_{1,\backslash}$.

\begin{figure*}[h]
\centering
\subfigure[]{
\begin{minipage}{7cm}
\centering
\includegraphics[width=0.8\textwidth]{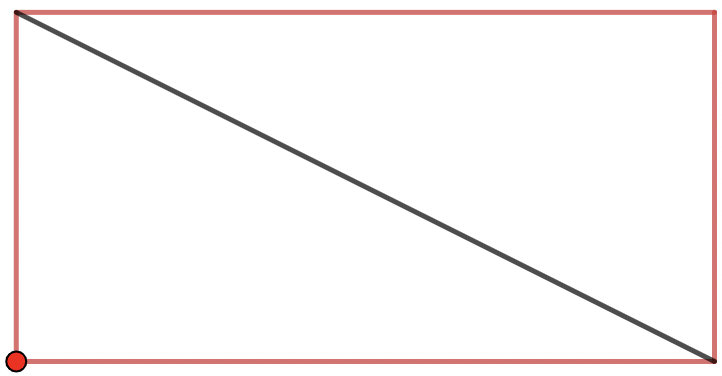}
\end{minipage}
}
\subfigure[]{
\begin{minipage}{7cm}
\centering
\includegraphics[width=0.7\textwidth]{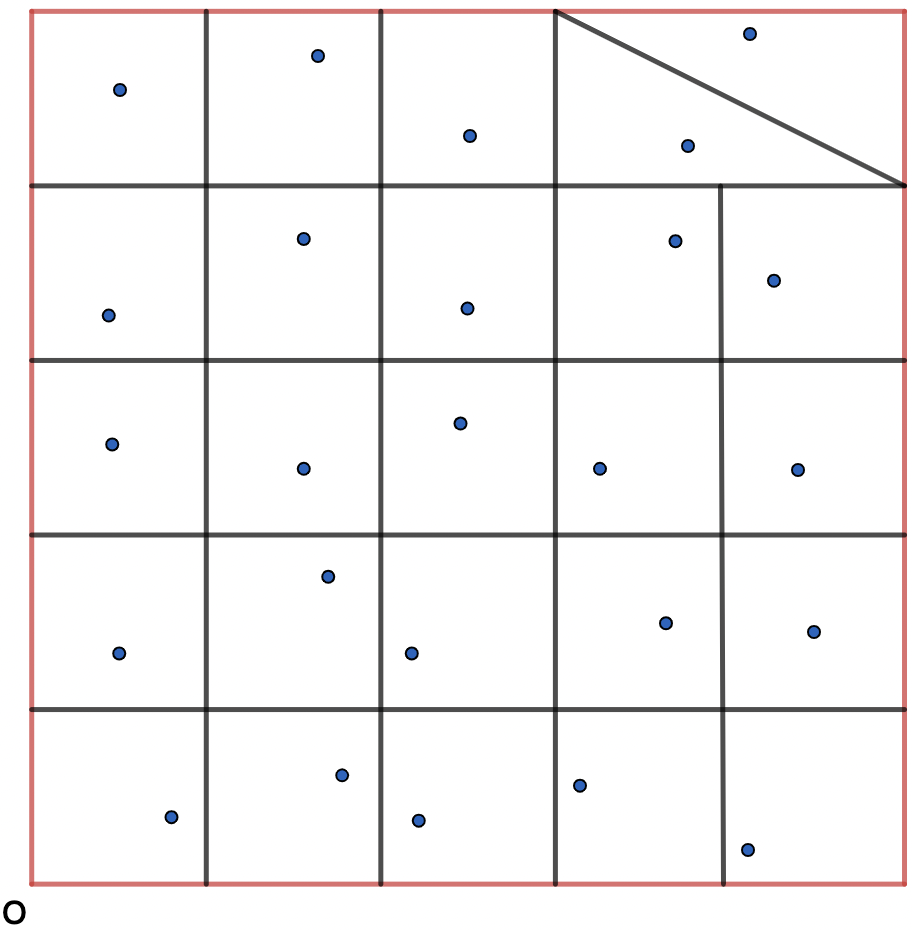}
\end{minipage}
}
\caption{\label{ss000} The designed partition model in \cite{KP}.}
\end{figure*}

\subsection{Class of partition model I}

For the merged rectangle $I$, we use a series of straight line partitions to divide the rectangle into two equal-volume parts, which will be converted to a one-parameter model if we set the angle between the dividing line and horizontal line across the center $\theta$, where we suppose $0\leq\theta\leq\frac{\pi}{2}$. From simple calculations, we can conclude the arbitrary straight line must pass through \textbf{the center of the rectangle}. For convenience of notation, we set this partition model $\Omega_{\sim}=(\Omega_{1,\sim},\Omega_{2,\sim},Q_3,\ldots,Q_{N})$ in two dimensional case.

\begin{figure}[H]
\centering
\includegraphics[width=0.30\textwidth]{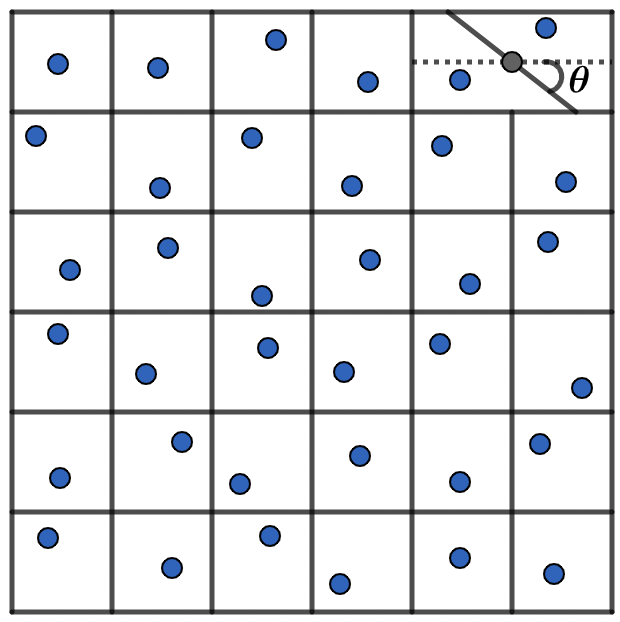}
\caption{\label{thtass00}A class of partitions for two dimensions}
\end{figure}

\subsection{Class of partition model II}

For the rectangle $$I=[a_1,a_1+2b]\times [a_2,a_2+b],$$where $a_1,a_2,b$ are three positive constants. A straight line partition is used to divide the rectangle into two parts if we set the straight line parallel to the diagonal of $I$, and the distance from the intersection point $q$ to the endpoint at the upper right corner of $I$ is $b\in (0,\frac{2}{m})$.

For convenience of notation, we set this partition model $$\Omega_{b,\sim}=(\Omega_{1,b,\sim},\Omega_{2,b,\sim},Q_3,\ldots,Q_{N}),$$ where $\Omega_{2,b,\sim}=I\setminus\Omega_{1,b,\sim}$.

\begin{figure}[H]
\centering
    \includegraphics[width=0.40\textwidth]{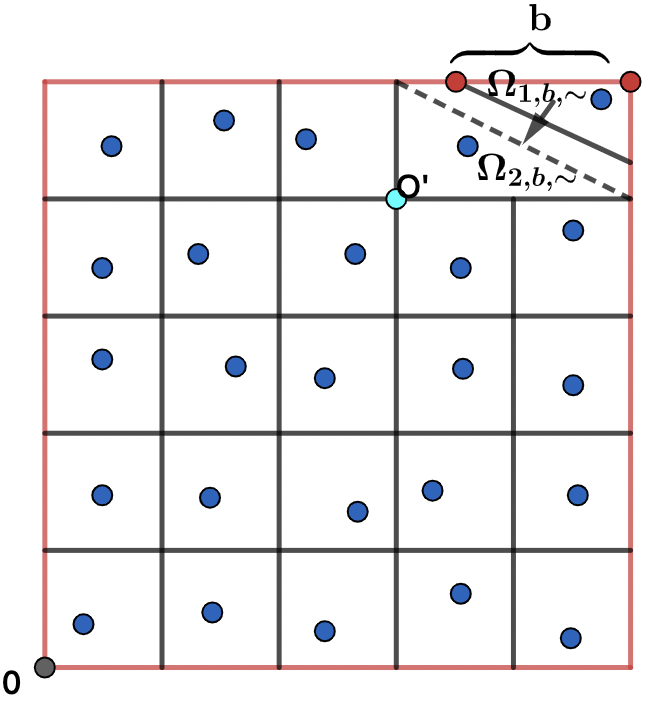}
\caption{\label{unpar} Uneqivolume partition}
\end{figure}

Now, consider $d-$dimensional cuboid
\begin{equation}\label{eq28}
    I_d=I\times\prod_{i=3}^{d}[a_i,a_i+b]
\end{equation} and its two partitions $\Omega'_{\sim}=(\Omega'_{1,\sim},\Omega'_{2,\sim})$ and $\Omega'_{b,\sim}=(\Omega'_{1,b,\sim},\Omega'_{2,b,\sim})$ into two closed, convex bodies with

\begin{equation}\label{o1blaksla}
    \Omega'_{1,\sim}= \Omega_{1,\sim}\times \prod_{i=3}^{d}[a_i,a_i+b],
\end{equation}
and

$$
\begin{aligned}
   \Omega'_{1,b,\sim}= \Omega_{1,b,\sim}\times \prod_{i=3}^{d}[a_i,a_i+b].
\end{aligned}
$$

We choose $a_1=\frac{m-2}{m}, a_2=\frac{m-1}{m}, b=\frac{1}{m}$ in $\Omega'_{1,\sim}$ and $\Omega'_{1,b,\sim}$, denoted by $\Omega^{*}_{1,\sim}$ and $\Omega^{*}_{1,b,\sim}$, then we obtain

\begin{equation}\label{omga1}
\Omega^{*}_{\sim}=(\Omega^{*}_{1,\sim},\Omega^{*}_{2,\sim},Q_3 \ldots,Q_{N}),
\end{equation}
and

\begin{equation}\label{omgaBsim}
\Omega^{*}_{b,\sim}=(\Omega^{*}_{1,b,\sim},\Omega^{*}_{2,b,\sim},Q_3 \ldots,Q_{N}).
\end{equation}

\section{Explicit Expected $L_2$-discrepancy for stratified random sampling formed by two classes of partitions}\label{sec3}
\subsection{Main results}
In this section, explicit $L_2-$discrepancy formulas are given for Class of partition model I and II.

\begin{thm}\label{expstarnp}
For partition $\Omega^{*}_{\theta,\sim}$ of $[0,1]^d$ and $m\ge 2, 0\leq \theta\leq \frac{\pi}{2}$, then

\begin{equation}\label{ednstarx}
\mathbb{E}L_2^2(D_N,P_{\Omega^{*}_{\theta,\sim}})= \frac{1}{m^{2d}}[(\frac{m-1}{2}+\frac{1}{2})^d-(\frac{m-1}{2}+\frac{1}{3})^d]+\frac{1}{m^{3d}}\cdot\frac{1}{3^{d}}\cdot P(\theta),
\end{equation} 
where

\begin{equation}\label{ptheta34}
     P(\theta)=\left\{
\begin{aligned}
&\frac{2}{5}tan^3\theta+\frac{6}{5}tan^2\theta-\frac{3tan\theta}{2}, \quad 0\leq\theta< arctan\frac{1}{2},
\\&-\frac{2}{5}, \quad  \theta=arctan\frac{1}{2},\\& -\frac{3}{8tan\theta}+\frac{3}{40tan^2\theta}+\frac{1}{160tan^3\theta}, \quad  arctan\frac{1}{2}<\theta\leq\frac{\pi}{2}.
\end{aligned}
\right.
\end{equation}
\end{thm}

\begin{figure}[H]
\centering
\includegraphics[width=0.50\textwidth]{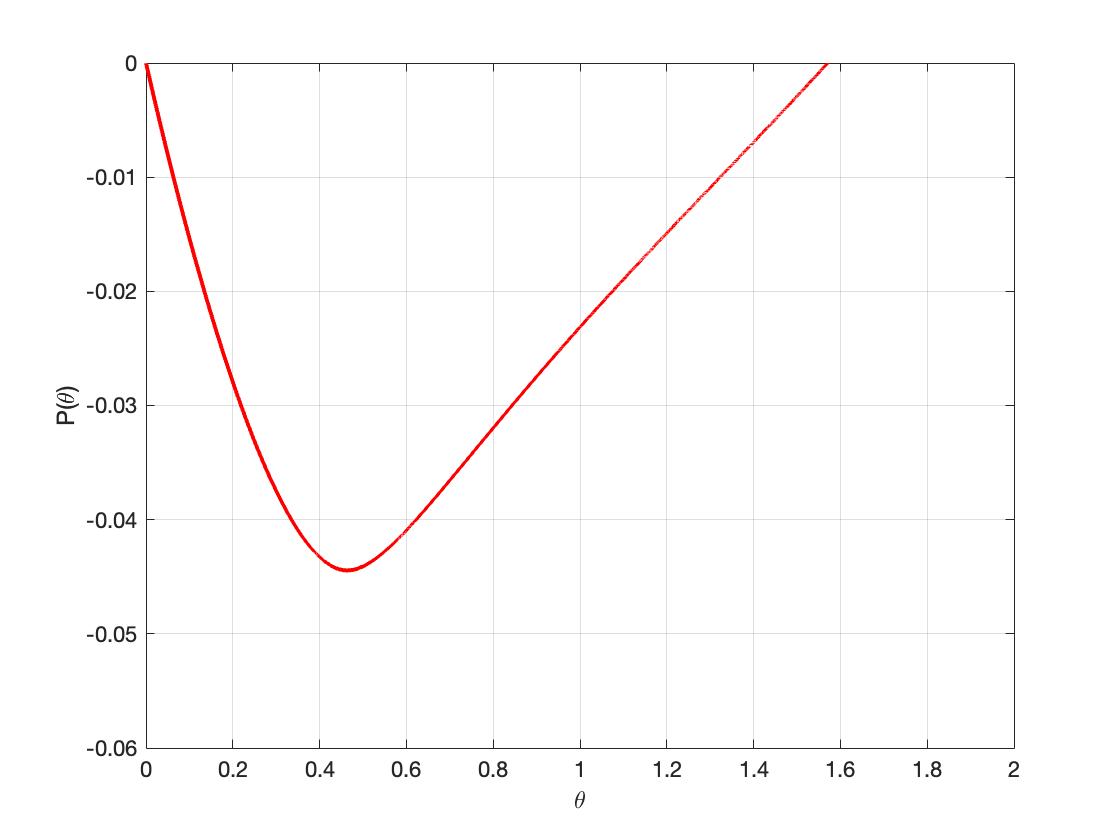}
\caption{\label{6p}$P(\theta)$ function}
\end{figure}

\begin{rem}
Noticing that in Theorem \ref{expstarnp}, $P(\theta)$ is a continuous function, decreases monotonically between $0$ and $arctan\frac{1}{2}$ and increases monotonically between $arctan\frac{1}{2}$ and $\frac{\pi}{2}$, see Figure \ref{6p}.  Choose parameter $\theta=\frac{\pi}{2}$ in Theorem \ref{expstarnp}, then we are back to the case of classical jittered sampling. 
Besides, the interesting thing is $\mathbb{E}L_2^2(D_N,P_{\Omega^{*}_{\theta,\sim}})$ is a continuous function of $\theta$, it dynamically establishes continuous dependency with partition parameters and provides ideas for continuous improvement of grid-based partition.\end{rem}

\begin{cor}
For partition $\Omega^{*}_{\theta,\sim}$ of $[0,1]^d$ and $m\ge 2, 0\leq \theta\leq \frac{\pi}{2}$, then

\begin{equation}
\mathbb{E}(L_2^2(D_N,P_{\Omega^{*}_{\theta,\sim}}))\leq \mathbb{E}L_{2}^2(D_{N},P_{\Omega^{*}_{|}}),
\end{equation}
where the equality holds if and only if $\theta=0$ and $\frac{\pi}{2}$.
\end{cor}

\begin{rem}
Noticing that $P(\theta)\leq 0$ and $P(\theta)=0$ if and only if $\theta=0$ and $\frac{\pi}{2}$. We already know it is shown in \cite{KP} the following conclusion
$$
\mathbb{E}(L_2^2(D_N,P_{\Omega^{*}_{\backslash}}))< \mathbb{E}L_{2}^2(D_{N},P_{\Omega^{*}_{|}}),
$$
holds, this is a counterexample that gives a theoretic conclusion, which proves the existence of a partition that with lower expected $L_2-$discrepancy than jittered sampling. This partition is a special case of the Class of partition model I, $\Omega^{*}_{\theta,\sim}$ for $\theta=arctan\frac{1}{2}$. Besides, this partition happens to achieve optimal expected $L_2-$discrepancy among the Class of partition model I, however, in \cite{KP}, the aim is to give a counterexample, that proves jittered sampling could not minimize expected $L_2-$discrepancy for all partition manners, the contribution of our Theorem \ref{expstarnp} is giving explicit expected $L_2-$discrepancy formulas under a class of equivolume partitions and improving the known expected $L_2-$discrepancy formula 
\begin{equation}\label{jittfor}
\mathbb{E}L_{2}^2(D_{N},P_{\Omega^*_|})=\frac{1}{m^{2d}}[(\frac{m-1}{2}+\frac{1}{2})^d-(\frac{m-1}{2}+\frac{1}{3})^d]
\end{equation}
recently derived in \cite{kirk2022expected}.
A smaller expected $L_2-$discrepancy formula gives better integration approximation error in a certain class of functional space if we consider some application aspects.
\end{rem}

\begin{thm}\label{th33}
For partition $\Omega^{*}_{b,\sim}$ of $[0,1]^d$ and $m\ge 2, b\in[\frac{3}{2m},\frac{2}{m}]$, then 

\begin{equation}\label{eq35}
\mathbb{E}L_{2}^2(D_{N},P_{\Omega^{*}_{b,\sim}})=\frac{1}{m^{2d}}[(\frac{m-1}{2}+\frac{1}{2})^d-(\frac{m-1}{2}+\frac{1}{3})^d]-\frac{P_0(b)}{2^d\cdot m^{3d}}-\frac{P_1(b)}{3^d\cdot m^{3d}},
\end{equation}
where
$$P_0(b)=\frac{8-m^2b^2}{3}-\frac{16}{24-3m^2b^2},$$

$$P_1(b)=\frac{m^4b^4}{40}+\frac{114m^2b^2}{40}+\frac{19}{5}-\frac{6m^3b^3-3m^5b^5+352}{40-5m^2b^2}.$$
\end{thm}

\begin{rem}
It can be easily analyzed that
$$\mathbb{E}L_{2}^2(D_{N},P_{\Omega^{*}_{b,\sim}})<\mathbb{E}L_2^2(D_N,P_{\Omega^{*}_{\theta,\sim}})$$
for any $\theta\in [0,\frac{\pi}{2}]$, thus it improves the expected $L_2-$discrepancy formula for jittered sampling, moreover, it attains better results than all cases of convex equivolume partitions as the class of partition model I. Therefore, expected $L_2-$discrepancy formulas under the class of partition I and II can be seen as the improvement of the Theorem 1 in \cite{kirk2022expected}, and the class of partition II can be seen as a class of partitions with lower expected $L_2-$discrepancy than jittered sampling and the partition model in \cite{KP}. The main contribution of Theorem \ref{th33} is giving explicit expected $L_2-$discrepancy formulas under a class of unequivolume partitions for $N=m^d$ in $d-$dimensional space, which provides a method to compute exact expected $L_2-$discrepancy formulas in $d-$dimensional space with $N=m^d$, by the way, it is proved that in the grid-like partition of $d$-dimensional space, the use of equivolume may not be optimal. This conclusion is a known result for two-dimensional space with two points, see \cite{KP2,PRS}, \cite{PRS} gives optimal partition manner(it happens to be unequivolume partition) for two-dimensional space with two points, for the case of $d-$dimensional space partitions with $N=m^d$, the partition model may be easy to extend, but the exact formula of expected $L_2-$discrepancy formula as \eqref{jittfor} under the optimal case remains unknown.\end{rem}

\subsection{Some discussions}

The model \cite{PRS} gives a family of unequivolume partitions in two dimensions with sampling number $N=2$, a certain type of symmetry along the $y = x$ diagonal is needed, and the partition is non-convex, expected $L_2-$discrepancy for $N=2, d=2$ attains optimal in these restrictions. It is also proved in \cite{KP2} some class of convex partitions could minimize the expected $L_2-$discrepancy for $d-$dimensions with $N=m^d$, thus an interesting question is could some class of convex unequivolume partitions achieve better expected $L_2-$discrepancy formulas than the family of non-convex partitions for $d$ with $N=m^d$? The family of partitions in \cite{PRS} is:

$$[0,1]^2=\Omega\cup ([0,1]^2\setminus \Omega),$$
where $\Omega=\{(x,y)\in[0,1]^2:y\leq g(x)\}$. If $g(x)$ satisfies a highly nonlinear integral equation, then the partition is optimal. This result can be extended to $d-$dimensions with $N=m^d$ as the partition model 2.4 in \cite{KP}, but it may be difficult to obtain an explicit expected $L_2-$discrepancy formula due to the difficulty in computing the solution of the integral equation that $g(x)$ has to satisfy. If we consider the convex partition, $g(x)$ will satisfy a linear equation, then an explicit formula is within reach. The optimal expected $L_2$-discrepancy formula may be obtained from the class of unequivolume partitions formed by a special class of linear equations, but we have not yet carried out the accurate calculation and comparison. Besides, since theorem \ref{th33} satisfies certain symmetry, it can be converted into a one-parameter model. Considering the partition class formed by general linear equations, the result should be a two-parameter model. As the equation satisfied by $g(x)$ becomes more and more complex, it is expected that the parameters contained in the $L_2-$discrepancy formula will become more complex, but it still has the value of numerical research, approximation error in \eqref{eq1d1} can still be accurately analyzed. 

\begin{figure*}[h]
\centering
\subfigure[Convex partitions]{
\begin{minipage}{7cm}
\centering
\includegraphics[width=0.6\textwidth]{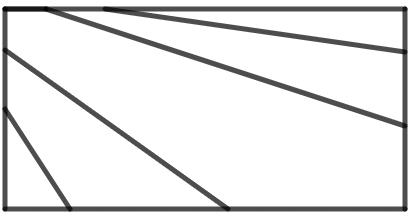}
\end{minipage}
}
\subfigure[Non-convex partitions]{
\begin{minipage}{7cm}
\centering
\includegraphics[width=0.7\textwidth]{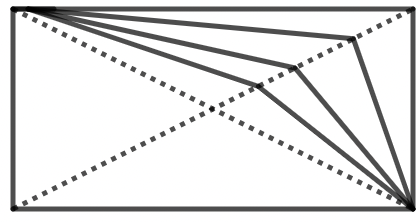}
\end{minipage}
}
\caption{\label{cornocpart} Some examples.}
\end{figure*}

For some missing convex or non-convex partitions, we can naturally provide the following two examples, see Figure \ref{cornocpart}, similar to model 2.4, they can be easily extended to $d-$dimensions with $N=m^d$. Expected $L_2-$discrepancy formulas are also easy to compute if we follow the proof process of Theorem \ref{expstarnp} and \ref{th33}, because their partition curve equations can always be explicitly displayed by some parameters.

\section{Proofs}\label{Prf}
\subsection{Proof of Theorem \ref{expstarnp}}

For equivolume partition $\mathbf{\Omega_{0,\sim}}=(\Omega_{1,\sim},\Omega_{2,\sim})$ of $I$(the same argument if we replace $\mathbf{\Omega_{0,\sim}}$ with $\mathbf{\Omega_{0,|}}$), from [Proposition $2$] in \cite{KP}, which is, for an equivolume partition $\hat{\Omega}=\{\hat{\Omega}_1, \hat{\Omega}_2, \ldots, \hat{\Omega}_N\}$ of a compact convex set $K\subset \mathbb{R}^{d}$ with $\lambda(K)>0$, $P_{\hat{\Omega}}$ is the corresponding stratified sampling set, then 

\begin{equation}\label{el22pom}
    \mathbb{E}L_2^2(D_N, P_{\hat{\Omega}})=\frac{1}{N^2\lambda(K)}\sum_{i=1}^{N}\int_{K}q_i(x)(1-q_i(x))dx,
\end{equation}
where 

\begin{equation}\label{qixx}
    q_i(x)=\frac{\lambda(\hat{\Omega}_i\cap[0,x])}{\lambda(\hat{\Omega}_i)}.
\end{equation}

Through simple derivation, it follows that
 \begin{equation}\label{18qi}
    \mathbb{E}L_2^2(D_N,P_{\mathbf{\Omega_{0,\sim}}})=\frac{1}{8}\sum_{i=1}^{2}\int_{I}\mathbf{q}_i(x)(1-\mathbf{q}_i(x))dx,
 \end{equation}
and

\begin{equation}\label{qix}
    \mathbf{q}_i(x)=\frac{\lambda(\Omega_{i,\sim}\cap [0,x])}{\lambda(\Omega_{i,\sim})}=\lambda(\Omega_{i,\sim}\cap [0,x]).
\end{equation}

Conclusion \eqref{18qi} is equivalent to the following

$$
    8\mathbb{E}L_2^2(D_N,P_{\mathbf{\Omega_{0,\sim}}})=1-\sum_{i=1}^{2}\int_{I}\mathbf{q}_i^{2}(x)dx.
$$

We first consider parameter $arctan\frac{1}{2}\leq\theta\leq\frac{\pi}{2}$, then we define the following two functions for simplicity of the expression.

$$
    F(\mathbf{x})=\frac{1}{2}\cdot[(x_1-1)tan\theta+x_2-\frac{1}{2}]\cdot[(x_1-1)+(x_2-\frac{1}{2})\cdot cot\theta],
$$
and

$$
    G(\mathbf{x})=x_1x_2-x_2-\frac{cot\theta}{2}x_2+\frac{1}{2}x_2^2\cdot cot\theta,
$$
where $\mathbf{x}=(x_1,x_2)$.

Furthermore, for $\mathbf{\Omega_{0,|}}=(\Omega_{1,|},\Omega_{2,|})$, \eqref{qix} implies

$$
    \mathbf{q}_{1,|}(\mathbf{x})=\left\{
\begin{aligned}
&x_1x_2, \mathbf{x}\in \Omega_{1,|}\\&
x_2, \mathbf{x}\in \Omega_{2,|},
\end{aligned}
\right.
$$
and 
$$
    \mathbf{q}_{2,|}(\mathbf{x})=\left\{
\begin{aligned}
&0, \mathbf{x}\in \Omega_{1,|}\\&
(x_1-1)x_2, \mathbf{x}\in \Omega_{2,|}.
\end{aligned}
\right.
$$

\begin{figure}[H]
\centering
\includegraphics[width=0.60\textwidth]{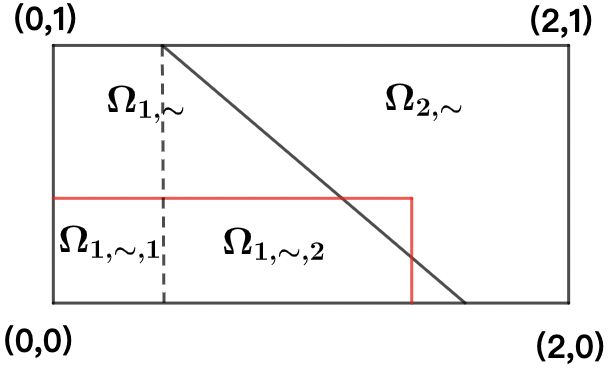}
\caption{Division of the integral region}\label{spt1}
\end{figure}

\begin{figure*}[h]
\centering
\subfigure[]{
\begin{minipage}{7cm}
\centering
\includegraphics[width=1.0\textwidth]{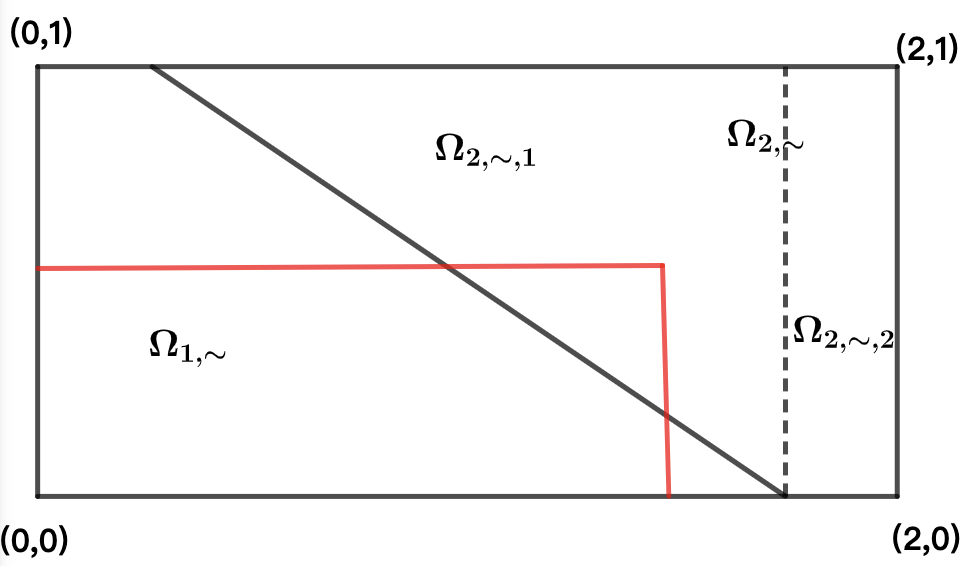}
\end{minipage}
}
\subfigure[]{
\begin{minipage}{7cm}
\centering
\includegraphics[width=1.0\textwidth]{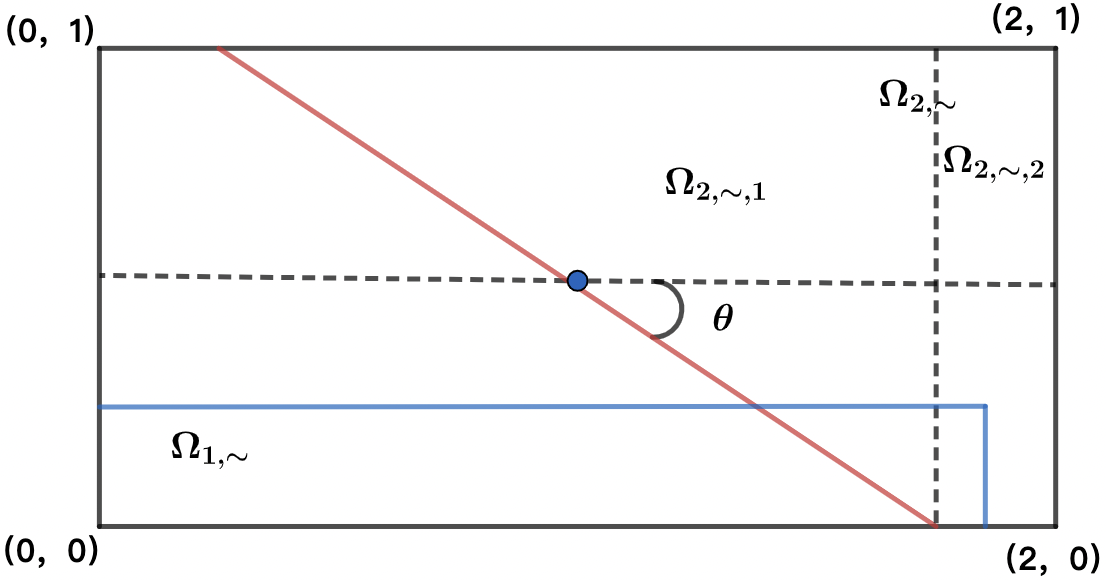}
\end{minipage}
}
\caption{\label{figure6} Division of the integral region.}
\end{figure*}

Besides, 

$$
    \mathbf{q}_{1,\sim}(\mathbf{x})=\left\{
\begin{aligned}
&x_1x_2, \mathbf{x}\in \Omega_{1,\sim},\\&
x_1x_2-F(\mathbf{x}), \mathbf{x}\in \Omega_{2,\sim,1},\\&x_1x_2-G(\mathbf{x}), \mathbf{x}\in \Omega_{2,\sim,2},
\end{aligned}
\right.
$$
and 
$$
    \mathbf{q}_{2,\sim}(\mathbf{x})=\left\{
\begin{aligned}
&0, \mathbf{x}\in \Omega_{1,\sim},\\&
F(\mathbf{x}), \mathbf{x}\in \Omega_{2,\sim,1},\\&G(\mathbf{x}), \mathbf{x}\in \Omega_{2,\sim,2},
\end{aligned}
\right.
$$
where $\Omega_{1,\sim}$, $\Omega_{2,\sim}$ denote subsets of partition $\mathbf{\Omega_{0,\sim}}$. In the following, we shall continue to divide subsets $\Omega_{1,\sim}=\{\Omega_{1,\sim,1},\Omega_{1,\sim,2}\}$ and $\Omega_{2,\sim}=\{\Omega_{2,\sim,1},\Omega_{2,\sim,2}\}$ to facilitate calculation. See Figures \ref{spt1} to \ref{figure6}.

Therefore, for $\theta=\frac{\pi}{2}$, we introduce two symbols $B_{1,|},B_{2,|}$ and have

\begin{equation}\label{b11}
    B_{1,|}=\int_{I}\mathbf{q}_{1,|}^2(\mathbf{x})d\mathbf{x}=\int_{\Omega_{1,|}}x_1^2x_2^2d\mathbf{x}+\int_{\Omega_{2,|}}x_2^2d\mathbf{x}=\frac{1}{9}+\frac{1}{3}=\frac{4}{9},
\end{equation}
and

\begin{equation}\label{b21}
  B_{2,|}=\int_{I}\mathbf{q}_{2,|}^2(\mathbf{x})d\mathbf{x}=\int_{\Omega_{2,|}}(x_1-1)^2x_2^2d\mathbf{x}=\frac{1}{9}.
\end{equation}

Thus,

\begin{equation}\label{8el2p1}
    8\mathbb{E}(L_2^2(P_{\mathbf{\Omega_{0,|}}}))=1-(B_{1,|}+B_{2,|})=\frac{4}{9}.
\end{equation}

Furthermore, we introduce $B_{1,\sim}$ and $B_{2,\sim}$, then

\begin{equation}\label{b1sim}
\begin{aligned}
    B_{1,\sim}=\int_{I}\mathbf{q}_{1,\sim}^2(\mathbf{x})d\mathbf{x}&=\int_{\Omega_{1,\sim}}x_1^2x_2^2d\mathbf{x}+\int_{\Omega_{2,\sim,1}}(x_1x_2-F(\mathbf{x}))^2d\mathbf{x}\\&+\int_{\Omega_{2,\sim,2}}(x_1x_2-G(\mathbf{x}))^2d\mathbf{x},
    \end{aligned}
\end{equation}
and

$$
  B_{2,\sim}=\int_{I}\mathbf{q}_{2,\sim}^2(\mathbf{x})d\mathbf{x}=\int_{\Omega_{2,\sim,1}}F^{2}(\mathbf{x})d\mathbf{x}+\int_{\Omega_{2,\sim,2}}G^{2}(\mathbf{x})d\mathbf{x}.$$
  
We divide our calculation into three steps. First, we compute $\int_{\Omega_{1,\sim}}x_1^2x_2^2d\mathbf{x}$, see Figure \ref{spt1} for illustration.

\begin{equation}\label{omega1sim1}
\begin{aligned}
    \int_{\Omega_{1,\sim,1}}x_1^2x_2^2d\mathbf{x}&=\int_{0}^{1-\frac{cot\theta}{2}}x_1^{2}dx_1\cdot \int_{0}^{1}x_2^{2}dx_2=\frac{(2-cot\theta)^{3}}{72}.
    \end{aligned}
\end{equation}

\begin{equation}\label{omega1sim2}
\begin{aligned}
    \int_{\Omega_{1,\sim,2}}x_1^2x_2^2d\mathbf{x}&=\int_{1-\frac{cot\theta}{2}}^{1+\frac{cot\theta}{2}}x_1^{2}dx_1\cdot \int_{0}^{(1-x_1)\cdot tan\theta+\frac{1}{2}}x_2^{2}dx_2
    \\&=\frac{60tan^{2}\theta-36tan\theta+7}{720tan^{3}\theta}.
    \end{aligned}
\end{equation}

Therefore, \eqref{omega1sim1} and \eqref{omega1sim2} imply

\begin{equation}\label{ome1sim}
\begin{aligned}
    \int_{\Omega_{1,\sim}}x_1^2x_2^2d\mathbf{x}&=\int_{\Omega_{1,\sim,1}}x_1^2x_2^2d\mathbf{x}+\int_{\Omega_{1,\sim,2}}x_1^2x_2^2d\mathbf{x}
    \\&=-\frac{1}{12tan\theta}+\frac{1}{30tan^2\theta}-\frac{1}{240tan^3\theta}+\frac{1}{9}.
\end{aligned}
\end{equation}

Second, we compute $\int_{\Omega_{2,\sim,1}}(x_1x_2-F(\mathbf{x}))^2d\mathbf{x}$ and $\int_{\Omega_{2,\sim,2}}(x_1x_2-G(\mathbf{x}))^2d\mathbf{x}$.

\begin{equation}\label{omega2sim1}
\begin{aligned}
    \int_{\Omega_{2,\sim,1}}(x_1x_2-F(\mathbf{x}))^2d\mathbf{x}&=\int_{1-\frac{cot\theta}{2}}^{1+\frac{cot\theta}{2}}\int_{(1-x_1)\cdot tan\theta+\frac{1}{2}}^{1}(x_1x_2-F(\mathbf{x}))^2dx_2dx_1\\&=\frac{180tan^{2}\theta-12tan\theta+5}{720tan^{3}\theta},
    \end{aligned}
\end{equation}

\begin{equation}\label{omega2sim2}
\begin{aligned}
    &\int_{\Omega_{2,\sim,2}}(x_1x_2-G(\mathbf{x}))^2d\mathbf{x}=\int_{1+\frac{cot\theta}{2}}^{2}\int_{0}^{1}(x_1x_2-G(\mathbf{x}))^2dx_2dx_1\\&=-\frac{cot^3\theta}{240}-\frac{cot^2\theta}{30}-\frac{cot\theta}{12}+\frac{1}{3}.
    \end{aligned}
\end{equation}

Thus, \eqref{omega2sim1} and \eqref{omega2sim2} imply 

\begin{equation}\label{ome2sim23}
\begin{aligned}
    &\int_{\Omega_{2,\sim,1}}(x_1x_2-F(\mathbf{x}))^2d\mathbf{x}+\int_{\Omega_{2,\sim,2}}(x_1x_2-G(\mathbf{x}))^2d\mathbf{x}\\&=\frac{1}{3}+\frac{1}{6tan\theta}-\frac{1}{20tan^2\theta}+\frac{1}{360tan^3\theta}.
\end{aligned}
\end{equation}

Combining \eqref{b1sim}, \eqref{ome1sim} and \eqref{ome2sim23}, we have

\begin{equation}\label{B1sim1}
\begin{aligned}
     B_{1,\sim}&=\int_{\Omega_{1,\sim}}x_1^2x_2^2d\mathbf{x}+\int_{\Omega_{2,\sim,1}}(x_1x_2-F(\mathbf{x}))^2d\mathbf{x}+\int_{\Omega_{2,\sim,2}}(x_1x_2-G(\mathbf{x}))^2d\mathbf{x}\\&=\frac{1}{12tan\theta}-\frac{1}{60tan^2\theta}-\frac{1}{720tan^3\theta}+\frac{4}{9}.
    \end{aligned}
\end{equation}

Third, we will compute $\int_{\Omega_{2,\sim,1}}F^{2}(\mathbf{x})d\mathbf{x}$ and $\int_{\Omega_{2,\sim,2}}G^2(\mathbf{x})d\mathbf{x}$ in the following.

In fact,

\begin{equation}\label{omega22sim1}
\begin{aligned}
    \int_{\Omega_{2,\sim,1}}F^{2}(\mathbf{x})d\mathbf{x}&=\int_{1-\frac{cot\theta}{2}}^{1+\frac{cot\theta}{2}}\int_{(1-x_1)\cdot tan\theta+\frac{1}{2}}^{1}F^{2}(\mathbf{x})dx_2dx_1\\&=\frac{1}{120tan^{3}\theta},
    \end{aligned}
\end{equation}

\begin{equation}\label{omega22sim2}
\begin{aligned}
    &\int_{\Omega_{2,\sim,2}}G^{2}(\mathbf{x})d\mathbf{x}=\int_{1+\frac{cot\theta}{2}}^{2}\int_{0}^{1}G^{2}(\mathbf{x})dx_2dx_1\\&=\frac{1}{9}-\frac{1}{24tan\theta}+\frac{1}{120tan^2\theta}-\frac{11}{1440tan^3\theta}.
    \end{aligned}
\end{equation}

Combining \eqref{omega22sim1} and \eqref{omega22sim2}, we have

\begin{equation}\label{ome2sim}
\begin{aligned}
    B_{2,\sim}&=\int_{\Omega_{2,\sim,1}}F^2(\mathbf{x})d\mathbf{x}+\int_{\Omega_{2,\sim,2}}G^2(\mathbf{x})d\mathbf{x}\\&=\frac{1}{9}-\frac{1}{24tan\theta}+\frac{1}{120tan^2\theta}+\frac{1}{1440tan^3\theta}.
\end{aligned}
\end{equation}

Thus, 

\begin{equation}\label{b1simb2sim}
   B_{1,\sim}+B_{2,\sim}=\frac{1}{24tan\theta}-\frac{1}{120tan^2\theta}-\frac{1}{1440tan^3\theta}+\frac{5}{9}.
\end{equation}

Therefore,

\begin{equation}\label{el2p}
\begin{aligned}
    8\mathbb{E}(L_2^2(P_{\mathbf{\Omega_{0,\sim}}}))&=1-(B_{1,\sim}+B_{2,\sim})
    \\&=-\frac{cot\theta}{24}+\frac{cot^2\theta}{120}+\frac{cot^3\theta}{1440}+\frac{4}{9},
    \end{aligned}
\end{equation}
where $arctan\frac{1}{2}\leq\theta<\frac{\pi}{2}$. 

For $\theta=\frac{\pi}{2}$, by \eqref{8el2p1} we have

\begin{equation}\label{el22149}
8\mathbb{E}(L_2^2(P_{\mathbf{\Omega_{0,|}}}))=\frac{4}{9}.
\end{equation}

\begin{figure*}[h]
\centering
\subfigure[]{
\begin{minipage}{7cm}
\centering
\includegraphics[width=1.0\textwidth]{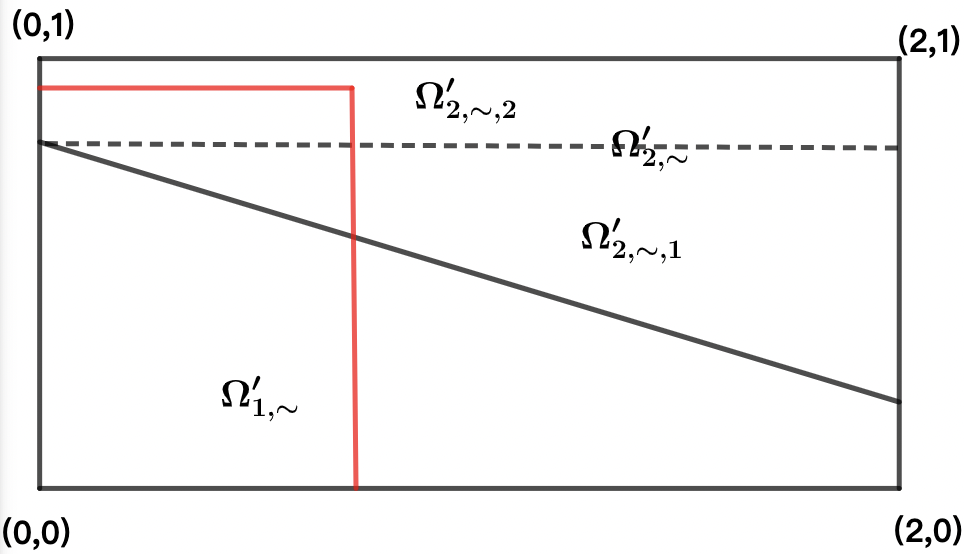}
\end{minipage}
}
\subfigure[]{
\begin{minipage}{7cm}
\centering
\includegraphics[width=1.0\textwidth]{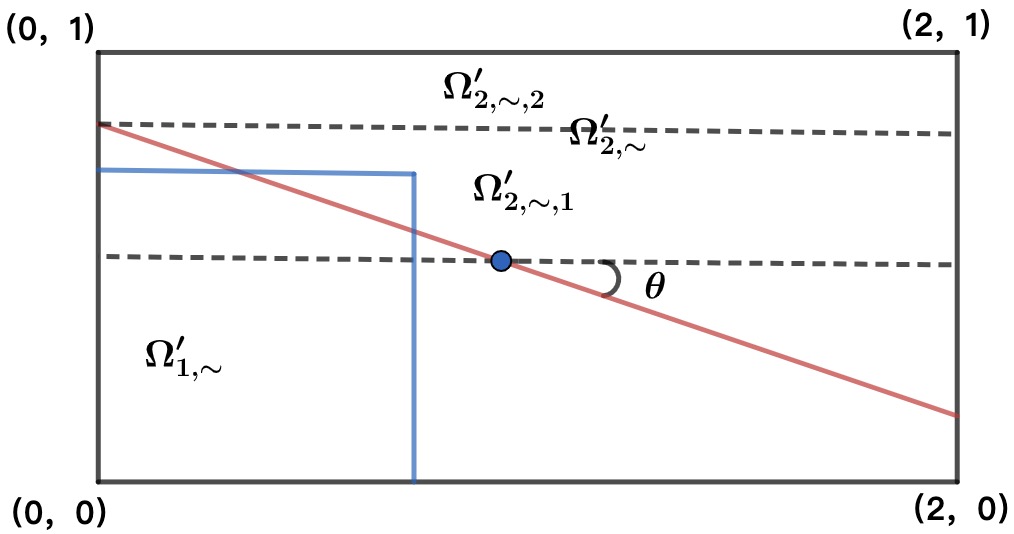}
\end{minipage}
}
\caption{\label{figure7} Division of the integral region.}
\end{figure*}

Considering the case $0\leq\theta<arctan\frac{1}{2}$, we denote the partition by $\Omega'_{\sim}=\{\Omega'_{1,\sim},\Omega'_{2,\sim}\}$, see Figure \ref{figure7}. Let

$$
    \mathbf{q}'_{1,\sim}(\mathbf{x})=\left\{
\begin{aligned}
&x_1x_2, \mathbf{x}\in \Omega'_{1,\sim},\\&
x_1x_2-H(\mathbf{x}), \mathbf{x}\in \Omega'_{2,\sim,1},\\& x_1x_2-J(\mathbf{x}), \mathbf{x}\in \Omega'_{2,\sim,2}.
\end{aligned}
\right.
$$
and 
$$
    \mathbf{q}'_{2,\sim}(\mathbf{x})=\left\{
\begin{aligned}
&0, \mathbf{x}\in \Omega'_{1,\sim},\\&
H(\mathbf{x}), \mathbf{x}\in \Omega'_{2,\sim,1},\\&
J(\mathbf{x}), \mathbf{x}\in \Omega'_{2,\sim,2},
\end{aligned}
\right.
$$
where
\begin{equation}
    H(x)=\frac{1}{2}\cdot[x_2-(1-x_1)tan\theta-\frac{1}{2}]\cdot[cot\theta\cdot x_2-1+x_1-\frac{1}{2}cot\theta],
\end{equation}
and

\begin{equation}
    J(x)=[x_2-tan\theta-\frac{1}{2}]\cdot x_1+\frac{1}{2}x_1^2\cdot tan\theta.
\end{equation}
Then we divide subsets $\Omega'_{1,\sim}=\{\Omega'_{1,\sim,1},\Omega'_{1,\sim,2}\}$ and $\Omega'_{2,\sim}=\{\Omega'_{2,\sim,1}, \Omega'_{2,\sim,2}\}$ to facilitate calculation. See Figure \ref{figure7}.

So 

\begin{equation}\label{eq424}
\begin{aligned}
    B'_{1,\sim}=\int_{I}\mathbf{q}_{1,\sim}^{'2}(\mathbf{x})d\mathbf{x}&=\int_{\Omega'_{1,\sim}}x_1^2x_2^2d\mathbf{x}+\int_{\Omega'_{2,\sim,1}}(x_1x_2-H(\mathbf{x}))^2d\mathbf{x}\\&+\int_{\Omega'_{2,\sim,2}}(x_1x_2-J(\mathbf{x}))^2d\mathbf{x},
    \end{aligned}
\end{equation}
and

$$
  B'_{2,\sim}=\int_{I}\mathbf{q}_{2,\sim}^{'2}(\mathbf{x})d\mathbf{x}=\int_{\Omega'_{2,\sim,1}}H^{2}(\mathbf{x})d\mathbf{x}+\int_{\Omega'_{2,\sim,2}}J^{2}(\mathbf{x})d\mathbf{x}.
$$

If we follow the calculation process of \eqref{omega1sim1}-\eqref{b1simb2sim}, then we obtain 

\begin{equation}
    B'_{1,\sim}=-\frac{4}{45}tan^3\theta-\frac{4}{15}tan^2\theta+\frac{tan\theta}{3}+\frac{4}{9},
\end{equation}
and

\begin{equation}\label{b2sim}
    B'_{2,\sim}=\frac{2}{45}tan^3\theta+\frac{2}{15}tan^2\theta-\frac{tan\theta}{6}+\frac{1}{9}.
\end{equation}

Thus,

\begin{equation}\label{b1simb2sim111}
   B'_{1,\sim}+B'_{2,\sim}=-\frac{2}{45}tan^3\theta-\frac{2}{15}tan^2\theta+\frac{tan\theta}{6}+\frac{5}{9}.
\end{equation}

Hence,

\begin{equation}\label{lygjd}
\begin{aligned}
    8\mathbb{E}(L_2^2(P_{\Omega'_{\sim}}))&=1-(B'_{1,\sim}+ B'_{2,\sim})\\&=\frac{4}{9}+\frac{2}{45}tan^3\theta+\frac{2}{15}tan^2\theta-\frac{tan\theta}{6},
    \end{aligned}
\end{equation}
where $0\leq\theta<arctan\frac{1}{2}$.

Combining with \eqref{el2p} and considering the translation and stretch of the rectangle $I=[0,2]\times[0,1]$ into $$[a_1,a_1+2b]\times [a_2,a_2+b],$$ we obtain

\begin{equation}\label{440comp28}
\mathbb{E}(L_2^2(P_{\Omega^{*}_{\sim}}))\leq\mathbb{E}(L_2^2(P_{\Omega^{*}_{|}})),
\end{equation}
where $a_1=\frac{m-2}{m},a_2=\frac{m-1}{m},b=\frac{1}{m}$. The equal sign of \eqref{440comp28} holds if and only if partition parameter $\theta=0,\frac{\pi}{2}$. Noting that conclusion \eqref{440comp28} is only for the two-dimensional case.

Next, we will give a proof of \eqref{440comp28} for $d-$dimensional case. We firstly prove the case $b=1$ and $(a_1,a_2,\ldots,a_d)=(0,0,\ldots,0).$ Let $I'_d=[0,2]\times[0,1]\times[0,1]^{d-2}$ and we denote partition manner of this special case $\Omega''_{\sim}=\{\Omega''_{1,\sim},\Omega''_{2,\sim}\}$.  

For $i=1,2$, we have

$$
    \mathbf{q}'_{i,\sim}(\mathbf{x})=\mathbf{q}_{i,\sim}(x_1,x_2)\cdot\prod_{j=3}^{d}x_j,
$$
where $\mathbf{q}'_{i,\sim}(\mathbf{x})$ is defined as \eqref{qix} for $\Omega''_{\sim}$.

Thus,

$$
    \int_{I'_d} \mathbf{q}'^{2}_{i,\sim}(\mathbf{x})d\mathbf{x}=B_{i,\sim}\cdot\int_{[0,1]^{d-2}}\prod_{j=3}^{d}x_j^2dx_3dx_4\ldots dx_d=\frac{1}{3^{d-2}}\cdot B_{i,\sim},
$$
where $B_{i,\sim},i=1,2$ have been calculated in \eqref{B1sim1} and \eqref{ome2sim} respectively.

As we have 

$$
    \int_{I'_d}\lambda([0,\mathbf{x}])d\mathbf{x}=\int_{[0,1]^{d-2}}\prod_{j=3}^{d}x_jdx_3dx_4\ldots dx_d=\frac{1}{2^{d-2}}.
$$

Then we obtain,

\begin{equation}\label{8l22omesim}
8\mathbb{E}(L_2^2(P_{\Omega''_{\sim}}))=\frac{1}{2^{d-2}}-\frac{1}{3^{d-2}}\cdot(B_{1,\sim}+B_{2,\sim}).
\end{equation}

Now, for $I_d$ in \eqref{eq28}, we define a vector

\begin{equation}
    \mathbf{a}=\{a_1,a_2,\ldots,a_d\}.
\end{equation}

We then prove \eqref{el22pom} is independent of $\mathbf{a}$. In $I_d$, we choose $\mathbf{a}=0$, set

\begin{equation}\label{id0433}
    I_{d}^{0}=[0,2b]\times [0,b]^{d-1},
\end{equation}
and
\begin{equation}\label{id0m433}
    I_{d,m}^{0}=[0,\frac{2}{m}]\times [0,\frac{1}{m}]^{d-1}.
\end{equation}

It suffices to show that

\begin{equation}\label{eq434}
     \frac{1}{N^2\lambda(I_d)}\sum_{i=1}^{N}\int_{I_d}q_i(x)(1-q_i(x))dx=\frac{1}{N^2\lambda(I_{d}^{0})}\sum_{i=1}^{N}\int_{I_{d}^{0}}q_i(x)(1-q_i(x))dx.
\end{equation}

We only consider $N=2$ in \eqref{eq434}, this is because we choose $K=I_d$ and $K=I_{d}^{0}$ in \eqref{el22pom} respectively. This means $I_d,I_{d}^{0}$ are divided into two equal volume parts respectively.

Let 

\begin{equation}\label{xiaiti}
    x_i-a_i=t_i,1\leq i\leq d.
\end{equation}

According to \eqref{qixx} and plugging \eqref{xiaiti} into the left side of \eqref{eq434}, the desired result is obtained.

From \eqref{el22pom} and let $K=[0,1]^d$, we have

\begin{equation}\label{eq437}
\begin{aligned}
    &\mathbb{E}L_2^2(P_{\Omega^{*}_{\sim}})-\mathbb{E}L_2^2(P_{\Omega^{*}_{|}})\\&=\frac{1}{N^2}\sum_{i=1}^{N}\int_{[0,1]^d}\tilde{q}_i(x)(1-\tilde{q}_i(x))dx-\frac{1}{N^2}\sum_{i=1}^{N}\int_{[0,1]^d}\bar{q}_i(x)(1-\bar{q}_i(x))dx,
\end{aligned}
\end{equation}
where 
$$ \tilde{q}_i(x)=\frac{\lambda(\Omega^{*}_{i,\sim}\cap[0,x])}{\lambda(\Omega^{*}_{i,\sim})}, \bar{q}_i(x)=\frac{\lambda(\Omega^{*}_{i,|}\cap[0,x])}{\lambda(\Omega^{*}_{i,|})}, i=1,2,$$
and 
$$\tilde{q}_i(x)=\bar{q}_i(x)=\frac{\lambda(Q_i\cap[0,x])}{\lambda(Q_i)}, i=3,4,\ldots,N.$$

Let $I_{d,m}^{0}=\{\Omega^{*}_{1,\sim},\Omega^{*}_{2,\sim}\}$, $I_{d,m}^{0}=\{\Omega^{*}_{1,|},\Omega^{*}_{2,|}\}$ denote two different partitions of $I_{d,m}^{0}$. It can easily be seen only $I_{d,m}^{0}$ contributes to the difference between two expected $L_2-$discrepancies, thus

\begin{equation}\label{eq438}
\begin{aligned}
    &\mathbb{E}L_2^2(P_{\Omega^{*}_{\sim}})-\mathbb{E}L_2^2(P_{\Omega^{*}_{|}})
    \\&=\frac{1}{N^2}\sum_{i=1}^{2}\int_{I_{d,m}^{0}}(\tilde{q}_i(x)-\bar{q}_i(x))dx+\frac{1}{N^2}\sum_{i=1}^{2}\int_{I_{d,m}^{0}}(\bar{q}^2_i(x)-\tilde{q}^2_i(x))dx\\&=\frac{1}{N}\sum_{i=1}^{2}\int_{I_{d,m}^{0}}(\lambda(\tilde{\Omega}_i\cap[0,x])-\lambda(\bar{\Omega}_i\cap[0,x]))dx\\&+\sum_{i=1}^{2}\int_{I_{d,m}^{0}}(\lambda^2(\bar{\Omega}_i\cap[0,x])-\lambda^2(\tilde{\Omega}_i\cap[0,x]))dx\\&=\frac{1}{N^3}\sum_{i=1}^{2}\int_{I'_d}(\lambda(\Omega''_{i,\sim}\cap[0,x])-\lambda(\Omega''_{i,|}\cap[0,x]))dx\\&+\frac{1}{N^3}\sum_{i=1}^{2}\int_{I'_d}(\lambda^2(\Omega''_{i,|}\cap[0,x])-\lambda^2(\Omega''_{i,\sim}\cap[0,x]))dx.
\end{aligned}
\end{equation}

Furthermore, employing \eqref{el22pom} again, we have

\begin{equation}\label{eq439}
    \begin{aligned}
    &\mathbb{E}(L_2^2(P_{\Omega''_{\sim}}))-\mathbb{E}(L_2^2(P_{\Omega''_{|}}))\\&=\frac{1}{8}\sum_{i=1}^{2}\int_{I'_d}\mathbf{q}'_{i,\sim}(\mathbf{x})(1-\mathbf{q}'_{i,\sim}(\mathbf{x}))dx-\frac{1}{8}\sum_{i=1}^{2}\int_{I'_d}\mathbf{q}'_{i,|}(\mathbf{x})(1-\mathbf{q}'_{i,|}(\mathbf{x}))dx\\&=\frac{1}{8}\sum_{i=1}^{2}\int_{I'_d}(\lambda(\Omega''_{i,\sim}\cap[0,x])-\lambda(\Omega''_{i,|}\cap[0,x]))dx\\&+\frac{1}{8}\sum_{i=1}^{2}\int_{I'_d}(\lambda^2(\Omega''_{i,|}\cap[0,x])-\lambda^2(\Omega''_{i,\sim}\cap[0,x]))dx.
    \end{aligned}
\end{equation}

Combining with \eqref{eq438} and \eqref{eq439}, we obtain

\begin{equation}\label{eq431}
\mathbb{E}L_2^2(P_{\Omega^{*}_{\sim}})-\mathbb{E}L_2^2(P_{\Omega^{*}_{|}})=\frac{1}{N^3}\cdot [8\mathbb{E}(L_2^2(P_{\Omega''_{\sim}}))-8\mathbb{E}(L_2^2(P_{\Omega''_{|}}))].
\end{equation}

Combining with \eqref{b1simb2sim}, \eqref{b1simb2sim111} and \eqref{8l22omesim}, the proof is completed.

\subsection{Proof of Theorem 3.5} According to the definition of $L_2$-discrepancy, for point set $P_{\Omega^{*}_{b,\sim}}=\{s_1,s_2,\ldots,s_N\}$, it is easy to see

\begin{equation}\label{eqpoasim}
\mathbb{E}L_{2}^2(D_{N},P_{\Omega^{*}_{b,\sim}})=\int_{P_{\Omega^{*}_{b,\sim}}}\int_{[0,1]^{d}}|\lambda([0,z))- \frac{1}{N}\sum_{i=1}^{N}\mathbf{1}_{[0,z)}(s_{i})|^{2}dzd\omega.
\end{equation}

Let $$I_{\sim}=\Omega^{*}_{1,b,\sim}\cup\Omega^{*}_{2,b,\sim}.$$

By \eqref{eqpoasim}, we obtain
\begin{equation}\label{dwdwftxk}
    \begin{aligned}\mathbb{E}L_{2}^2(D_{N},P_{\Omega^{*}_{b,\sim}})&=\int_{P_{\Omega^{*}_{b,\sim}}}\int_{[0,1]^{d}}|\lambda([0,z))- \frac{1}{N}\sum_{i=1}^{N}\mathbf{1}_{[0,z)}(s_{i})|^{2}dzd\omega\\&=\int_{P_{\Omega^{*}_{b,\sim}}}\int_{[0,1]^d \setminus I_{\sim}}|\lambda([0,z))- \frac{1}{N}\sum_{i=1}^{N}\mathbf{1}_{[0,z)}(s_{i})|^{2}dzd\omega\\&+\int_{P_{\Omega^{*}_{b,\sim}}}\int_{I_{\sim}}|\lambda([0,z))- \frac{1}{N}\sum_{i=1}^{N}\mathbf{1}_{[0,z)}(s_{i})|^{2}dzd\omega.
 \end{aligned}
\end{equation}

\begin{figure}[H]
\centering
\includegraphics[width=0.40\textwidth]{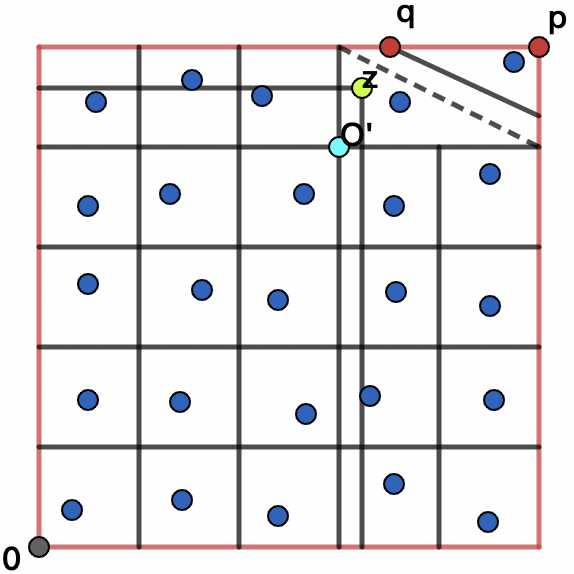}
\caption{}
\end{figure}

First, we focus on $I_{\sim}$, it can easily be seen

\begin{equation}\label{eqgsxk}
    \begin{aligned}&\int_{P_{\Omega^{*}_{b,\sim}}}\int_{I_{\sim}}|\lambda([0,z))- \frac{1}{N}\sum_{i=1}^{N}\mathbf{1}_{[0,z)}(s_{i})|^{2}dzd\omega\\&=\int_{P_{\Omega^{*}_{b,\sim}}}\int_{I_{\sim}}|\lambda([0,z)\setminus[O',z))+\lambda([O',z))- \frac{1}{N}\sum_{i=1}^{N}\mathbf{1}_{[0,z)\setminus[O',z)}(s_{i})- \frac{1}{N}\sum_{i=1}^{N}\mathbf{1}_{[O',z)}(s_{i})|^{2} dzd\omega.
    \end{aligned}
\end{equation}

Let $\Omega_z=[0,z)\setminus[O',z)$, if we divide $I_{\sim}$ into three parts, then \eqref{eqgsxk} implies

\begin{equation*}
 \begin{aligned}&\int_{P_{\Omega^{*}_{b,\sim}}}\int_{I_{\sim}}|\lambda([0,z))- \frac{1}{N}\sum_{i=1}^{N}\mathbf{1}_{[0,z)}(s_{i})|^{2}dzd\omega\\&=\int_{P_{\Omega^{*}_{b,\sim}}}\int_{I+II+III}|\lambda(\Omega_z)+\lambda([O',z))- \frac{1}{N}\sum_{i=1}^{N}\mathbf{1}_{\Omega_z}(s_{i})- \frac{1}{N}\sum_{i=1}^{N}\mathbf{1}_{[O',z)}(s_{i})|^{2}dzd\omega.
    \end{aligned}
\end{equation*}

\begin{figure}[H]
\centering
\includegraphics[width=0.40\textwidth]{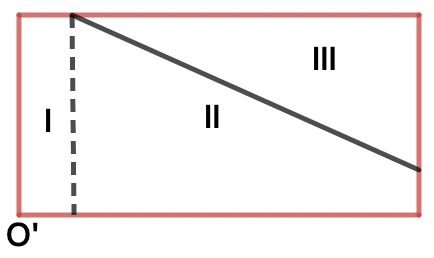}
\caption{Division of the integral region I}\label{sptI}
\end{figure}

We divide the computation process into two steps.

\textbf{The first step}, we compute

\begin{equation}\label{hfcwgqyfb}
\begin{aligned}
&\int_{P_{\Omega^{*}_{b,\sim}}}\int_{I+II}|\lambda(\Omega_z)+\lambda([O',z))- \frac{1}{N}\sum_{i=1}^{N}\mathbf{1}_{\Omega_z}(s_{i})- \frac{1}{N}\sum_{i=1}^{N}\mathbf{1}_{[O',z)}(s_{i})|^{2}dz d\omega\\&=\int_{I+II}\int_{P_{\Omega^{*}_{b,\sim}}}|\lambda(\Omega_z)+\lambda([O',z))- \frac{1}{N}\sum_{i=1}^{N}\mathbf{1}_{\Omega_z}(s_{i})- \frac{1}{N}\sum_{i=1}^{N}\mathbf{1}_{[O',z)}(s_{i})|^{2}
      d\omega dz.
    \end{aligned}
\end{equation}

Since

\begin{equation}\label{tjy1}
   \mathbb{E}(\frac{1}{N}\sum_{i=1}^{N}\mathbf{1}_{\Omega_z}(s_{i}))=\lambda(\Omega_z),
\end{equation}
and

\begin{equation}\label{tjy2}
\mathbb{E}(\frac{1}{N}\sum_{i=1}^{N}\mathbf{1}_{[O',z)}(s_{i}))=\frac{4m^{d-2}}{8m^{d-2}-Nb^2}\cdot\lambda([O',z)).
\end{equation}

If we set 

$$\mathbb{P}(s_i\in V_i)=\frac{\lambda(V_i)}{\lambda(\Omega^{*}_{i,b,\sim})}$$ for $i=1,2,$ and 

$$\mathbb{P}(s_i\in U_i)=\frac{\lambda(U_i)}{\lambda(Q_i)}$$ for $i=3,4,\ldots,N.$

Then, \eqref{hfcwgqyfb} can be converted to 

\begin{equation}\label{befcaozbf}
\begin{aligned}
&\int_{I+II}\int_{P_{\Omega^{*}_{b,\sim}}}| \mathbb{E}(\frac{1}{N}\sum_{i=1}^{N}\mathbf{1}_{\Omega_z}(s_{i}))+\mathbb{E}(\frac{1}{N}\sum_{i=1}^{N}\mathbf{1}_{[O',z)}(s_{i}))+(1-\frac{4m^{d-2}}{8m^{d-2}-Nb^2})\cdot\lambda([O',z))\\&- \frac{1}{N}\sum_{i=1}^{N}\mathbf{1}_{\Omega_z}(s_{i})- \frac{1}{N}\sum_{i=1}^{N}\mathbf{1}_{[O',z)}(s_{i})|^{2}
      d\omega dz\\&=\int_{I+II}\text{Var}( \frac{1}{N}\sum_{i=1}^{N}\mathbf{1}_{[0,z)\setminus [O',z)}(s_{i}))dz+\int_{I+II}\text{Var}( \frac{1}{N}\sum_{i=1}^{N}\mathbf{1}_{[O',z)}(s_{i}))\\&+|(1-\frac{4m^{d-2}}{8m^{d-2}-Nb^2})\cdot\lambda([O',z))|^2dz.
\end{aligned}
\end{equation}

\begin{equation*}
\text{Var}( \frac{1}{N}\sum_{i=1}^{N}\mathbf{1}_{[O',z)}(s_{i}))=\frac{1}{N^2}\cdot\frac{\lambda([O',z)])}{\lambda(I+II)}\cdot(1-\frac{\lambda([O',z)])}{\lambda(I+II)}).
\end{equation*}

\begin{equation*}
\lambda(I+II)=\frac{2}{N}-\frac{b^2}{4}\cdot\frac{1}{m^{d-2}}.
\end{equation*}

Thus,

\begin{equation*}
\begin{aligned}
&\text{Var}( \frac{1}{N}\sum_{i=1}^{N}\mathbf{1}_{[O',z)}(s_{i}))+|(1-\frac{4m^{d-2}}{8m^{d-2}-Nb^2})\cdot\lambda([O',z))|^2\\&=\frac{1}{N}\cdot \frac{4}{8-m^2b^2}\cdot \lambda([O',z)])+(1-\frac{8}{8-m^2b^2})\cdot\lambda^2([O',z)]).
\end{aligned}
\end{equation*}

The equation of the dividing line is the following

\begin{equation*}
Z_2=-\frac{1}{2}Z_1+\frac{3}{2}-\frac{b}{2}.
\end{equation*}

Besides, set $t=bm$, thus, 

\begin{equation*}
\begin{aligned}
 \int_{I}\lambda([O',z))&=\int_{1-\frac{2}{m}}^{1-b}\int_{1-\frac{1}{m}}^{1}(Z_1-1+\frac{2}{m})(Z_2-1+\frac{1}{m})dZ_1dZ_2\cdot\\&\int_{1-\frac{1}{m}}^{1}\int_{1-\frac{1}{m}}^{1}\ldots \int_{1-\frac{1}{m}}^{1}\prod_{i=3}^{d}(Z_i-1+\frac{1}{m})\\&=(t^2-4t+4)\cdot (\frac{1}{2m^2})^{d}.
 \end{aligned}
\end{equation*}

\begin{equation*}
\begin{aligned}
 \int_{II}\lambda([O',z))&=\int_{1-b}^{1}\int_{1-\frac{1}{m}}^{-\frac{Z_1}{2}+\frac{3}{2}-\frac{b}{2}}(Z_1-1+\frac{2}{m})(Z_2-1+\frac{1}{m})dZ_1dZ_2\cdot\\&\int_{1-\frac{1}{m}}^{1}\int_{1-\frac{1}{m}}^{1}\ldots \int_{1-\frac{1}{m}}^{1}\prod_{i=3}^{d}(Z_i-1+\frac{1}{m})\\&=(-\frac{t^4}{24}+\frac{2t^3}{3}-3t^2+4t)\cdot (\frac{1}{2m^2})^{d}.
 \end{aligned}
\end{equation*}

\begin{equation*}
\begin{aligned}
 \int_{I}\lambda^2([O',z))&=\int_{1-\frac{2}{m}}^{1-b}\int_{1-\frac{1}{m}}^{1}(Z_1-1+\frac{2}{m})^2(Z_2-1+\frac{1}{m})^2dZ_1dZ_2\cdot\\&\int_{1-\frac{1}{m}}^{1}\int_{1-\frac{1}{m}}^{1}\ldots \int_{1-\frac{1}{m}}^{1}(\prod_{i=3}^{d}(Z_i-1+\frac{1}{m}))^2\\&=(-t^3+6t^2-12t+8)\cdot (\frac{1}{3m^3})^{d}.
 \end{aligned}
\end{equation*}

\begin{equation*}
\begin{aligned}
 \int_{II}\lambda^2([O',z))&=\int_{1-b}^{1}\int_{1-\frac{1}{m}}^{-\frac{Z_1}{2}+\frac{3}{2}-\frac{b}{2}}(Z_1-1+\frac{2}{m})^2(Z_2-1+\frac{1}{m})^2dZ_1dZ_2\cdot\\&\int_{1-\frac{1}{m}}^{1}\int_{1-\frac{1}{m}}^{1}\ldots \int_{1-\frac{1}{m}}^{1}\prod_{i=3}^{d}(Z_i-1+\frac{1}{m})\\&=(-\frac{t^6}{160}+\frac{3t^5}{20}-\frac{3t^4}{2}+7t^3-15t^2+12t)\cdot (\frac{1}{3m^3})^{d}.
 \end{aligned}
\end{equation*}

\textbf{The second step}, for area $z\in$III, we have,

$$\lambda(\Omega_{2,1})=\frac{1}{4}(Z_1+2Z_2+b-3)^2\cdot\prod_{i=3}^{d}(Z_i-1+\frac{1}{m}),$$

and

$$\lambda(\Omega_{2,2})=(Z_1-1+\frac{2}{m})(Z_2-1+\frac{1}{m})\cdot\prod_{i=3}^{d}(Z_i-1+\frac{1}{m})-\lambda(\Omega_{2,1}).$$

\begin{figure}[H]
\centering
\includegraphics[width=0.40\textwidth]{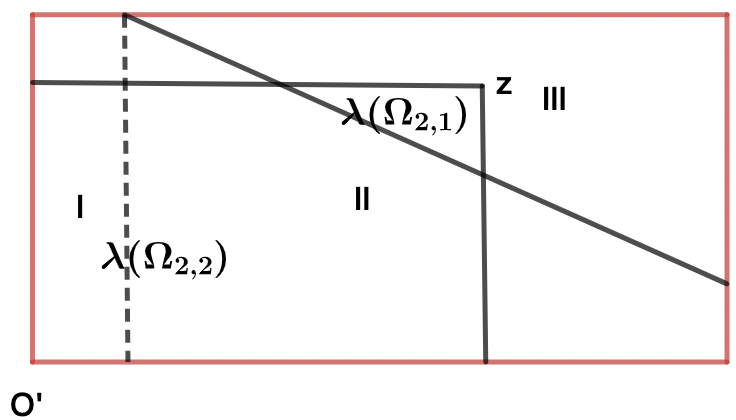}
\caption{Division of the integral region II}\label{sptI}
\end{figure}

Furthermore, we have

\begin{equation}
\begin{aligned}
&\int_{P_{\Omega^{*}_{b,\sim}}}\int_{III}|\lambda(\Omega_z)+\lambda([O',z))- \frac{1}{N}\sum_{i=1}^{N}\mathbf{1}_{\Omega_z}(s_{i})- \frac{1}{N}\sum_{i=1}^{N}\mathbf{1}_{[O',z)}(s_{i})|^{2}dz d\omega\\&=\int_{III}\int_{P_{\Omega^{*}_{b,\sim}}}|\lambda(\Omega_z)+\lambda([O',z))- \frac{1}{N}\sum_{i=1}^{N}\mathbf{1}_{\Omega_z}(s_{i})- \frac{1}{N}\sum_{i=1}^{N}\mathbf{1}_{[O',z)}(s_{i})|^{2}
      d\omega dz\\&\int_{III}\text{Var}( \frac{1}{N}\sum_{i=1}^{N}\mathbf{1}_{[0,z)\setminus [O',z)}(s_{i}))dz+\int_{III}\text{Var}( \frac{1}{N}\sum_{i=1}^{N}\mathbf{1}_{[O',z)}(s_{i}))\\&+|(1-\frac{4m^{d-2}}{Nb^2})\cdot\lambda(\Omega_{2,1})+(1-\frac{4m^{d-2}}{8m^{d-2}-Nb^2})\cdot\lambda(\Omega_{2,2})|^2dz.
    \end{aligned}
\end{equation}

Besides,
\begin{equation*}
\begin{aligned}
&\text{Var}( \frac{1}{N}\sum_{i=1}^{N}\mathbf{1}_{[O',z)}(s_{i}))+|(1-\frac{4m^{d-2}}{Nb^2})\cdot\lambda(\Omega_{2,1})+(1-\frac{4m^{d-2}}{8m^{d-2}-Nb^2})\cdot\lambda(\Omega_{2,2})|^2\\&=\text{Var}( \frac{1}{N}\sum_{i=1}^{N}\mathbf{1}_{[O',z)}(s_{i}))+|(1-\frac{4m^{d-2}}{Nb^2})\cdot\lambda(\Omega_{2,1})|^2+|(1-\frac{4m^{d-2}}{8m^{d-2}-Nb^2})\cdot\lambda(\Omega_{2,2})|^2\\&+2((1-\frac{4m^{d-2}}{Nb^2})\cdot\lambda(\Omega_{2,1}))((1-\frac{4m^{d-2}}{8m^{d-2}-Nb^2})\cdot\lambda(\Omega_{2,2}))\\&=\frac{4}{Nm^2b^2}\lambda(\Omega_{2,1})+\frac{4}{8N-Nm^2b^2}\lambda(\Omega_{2,2})+(1-\frac{8}{m^2b^2})\lambda^2(\Omega_{2,1})+(1-\frac{8}{8-m^2b^2})\lambda^2(\Omega_{2,2})\\&+2((1-\frac{4m^{d-2}}{Nb^2})\cdot\lambda(\Omega_{2,1}))((1-\frac{4m^{d-2}}{8m^{d-2}-Nb^2})\cdot\lambda(\Omega_{2,2})).
\end{aligned}
\end{equation*}

Moreover, we have,
\begin{equation*}
\begin{aligned}
&\int_{III}\lambda(\Omega_{2,2})dZ_1dZ_2\ldots Z_d\\&=\int_{1-b}^{1}\int_{-\frac{1}{2}Z_1+\frac{3}{2}-\frac{b}{2}}^{1}\int_{1-\frac{1}{m}}^{1}\ldots\int_{1-\frac{1}{m}}^{1}\lambda(\Omega_{2,2})dZ_1dZ_2\ldots Z_d\\&=\frac{-2t^3+6t^2}{3}(\frac{1}{2m^2})^{d},
\end{aligned}
\end{equation*}

\begin{equation*}
\begin{aligned}
&\int_{III}\lambda^2(\Omega_{2,2})dZ_1dZ_2\ldots Z_d\\&=\int_{1-b}^{1}\int_{-\frac{1}{2}Z_1+\frac{3}{2}-\frac{b}{2}}^{1}\int_{1-\frac{1}{m}}^{1}\ldots\int_{1-\frac{1}{m}}^{1}\lambda^2(\Omega_{2,2})dZ_1dZ_2\ldots Z_d\\&=(\frac{t^6}{80}-\frac{9t^5}{120}+\frac{9t^4}{8}-\frac{18t^3}{3}+9t^2)(\frac{1}{3m^3})^{d},
\end{aligned}
\end{equation*}

\begin{equation*}
\begin{aligned}
&\int_{III}\lambda(\Omega_{2,1})dZ_1dZ_2\ldots Z_d\\&=\int_{1-b}^{1}\int_{-\frac{1}{2}Z_1+\frac{3}{2}-\frac{b}{2}}^{1}\int_{1-\frac{1}{m}}^{1}\ldots\int_{1-\frac{1}{m}}^{1}\lambda(\Omega_{2,1})dZ_1dZ_2\ldots Z_d\\&=\frac{t^4}{24}(\frac{1}{2m^2})^{d},
\end{aligned}
\end{equation*}

\begin{equation*}
\begin{aligned}
&\int_{III}\lambda^2(\Omega_{2,1})dZ_1dZ_2\ldots Z_d\\&=\int_{1-b}^{1}\int_{-\frac{1}{2}Z_1+\frac{3}{2}-\frac{b}{2}}^{1}\int_{1-\frac{1}{m}}^{1}\ldots\int_{1-\frac{1}{m}}^{1}\lambda^2(\Omega_{2,1})dZ_1dZ_2\ldots Z_d\\&=\frac{9t^6}{960}(\frac{1}{3m^3})^{d},
\end{aligned}
\end{equation*}

\begin{equation*}
\begin{aligned}
&\int_{III}\lambda(\Omega_{2,1})\lambda(\Omega_{2,2}) dZ_1dZ_2\ldots Z_d\\&=\int_{1-b}^{1}\int_{-\frac{1}{2}Z_1+\frac{3}{2}-\frac{b}{2}}^{1}\int_{1-\frac{1}{m}}^{1}\ldots\int_{1-\frac{1}{m}}^{1}\lambda(\Omega_{2,1})\lambda(\Omega_{2,2})dZ_1dZ_2\ldots Z_d\\&=(-\frac{9t^6}{1152}-\frac{9t^5}{240}+\frac{9t^4}{48})(\frac{1}{3m^3})^{d}.
\end{aligned}
\end{equation*}

Therefore, we have

\begin{equation}\label{asimys2}
    \begin{aligned}\mathbb{E}L_{2}^2(D_{N},P_{\Omega^{*}_{b,\sim}})&=\int_{P_{\Omega^{*}_{b,\sim}}}\int_{[0,1]^{d}}|\lambda([0,z))- \frac{1}{N}\sum_{i=1}^{N}\mathbf{1}_{[0,z)}(s_{i})|^{2}dzd\omega\\&=\int_{P_{\Omega^{*}_{b,\sim}}}\int_{[0,1]^d \setminus I_{\sim}}|\lambda([0,z))- \frac{1}{N}\sum_{i=1}^{N}\mathbf{1}_{[0,z)}(s_{i})|^{2}dzd\omega\\&+\int_{P_{\Omega^{*}_{b,\sim}}}\int_{I_{\sim}}|\lambda([0,z))- \frac{1}{N}\sum_{i=1}^{N}\mathbf{1}_{[0,z)}(s_{i})|^{2}dzd\omega\\&=\int_{P_{\Omega^{*}_{b,\sim}}}\int_{[0,1]^d \setminus I_{\sim}}|\lambda([0,z))- \frac{1}{N}\sum_{i=1}^{N}\mathbf{1}_{[0,z)}(s_{i})|^{2}dzd\omega\\&+\int_{I_{\sim}}\text{Var}( \frac{1}{N}\sum_{i=1}^{N}\mathbf{1}_{[0,z)\setminus [O',z)}(s_{i})) dz+[(-\frac{t^4}{24}+4)\cdot \frac{4}{8-t^2}+\frac{t^2}{6}]\cdot (\frac{1}{2m^3})^d\\&+[(1-\frac{8}{t^2})\cdot\frac{9t^6}{960}+(1-\frac{8}{8-t^2})\cdot(\frac{t^6}{160}+\frac{9t^5}{120}-\frac{3t^4}{8}+8)\\&+2(1-\frac{4}{t^2})(1-\frac{4}{8-t^2})\cdot(-\frac{9t^6}{1152}-\frac{9t^5}{240}+\frac{9t^4}{48})]\cdot (\frac{1}{3m^3})^{d}.
 \end{aligned}
\end{equation}

According to $t=mb$, and let

$$P_2(b)=\frac{m^2b^2}{3}+\frac{16}{24-3m^2b^2}+\frac{4}{3},$$

$$P_3(b)=-\frac{m^4b^4}{40}-\frac{114m^2b^2}{40}-\frac{352}{40}+\frac{6m^3b^3-3m^5b^5+352}{40-5m^2b^2}.$$

We obtain

\begin{equation}\label{410}
\begin{aligned}\mathbb{E}L_{2}^2(D_{N},P_{\Omega^{*}_{b,\sim}})&=\int_{P_{\Omega^{*}_{b,\sim}}}\int_{[0,1]^d \setminus I_{\sim}}|\lambda([0,z))- \frac{1}{N}\sum_{i=1}^{N}\mathbf{1}_{[0,z)}(s_{i})|^{2}dzd\omega\\&+\int_{I_{\sim}}\text{Var}( \frac{1}{N}\sum_{i=1}^{N}\mathbf{1}_{[0,z)\setminus [O',z)}(s_{i})) dz+P_2(b)\cdot (\frac{1}{2m^3})^d+P_3(b)\cdot (\frac{1}{3m^3})^{d}.
\end{aligned}
\end{equation}

For jittered grid area $[0,1]^d \setminus I_{\sim}$ and $[0,z)\setminus [O',z)$, it is clear that

\begin{equation}\label{J1}
\begin{aligned}
&\int_{P_{\Omega^{*}_{b,\sim}}}\int_{[0,1]^d \setminus I_{\sim}}|\lambda([0,z))- \frac{1}{N}\sum_{i=1}^{N}\mathbf{1}_{[0,z)}(s_{i})|^{2}dzd\omega+\int_{I_{\sim}}\text{Var}( \frac{1}{N}\sum_{i=1}^{N}\mathbf{1}_{[0,z)\setminus [O',z)}(s_{i})) dz\\&=\int_{P_{\Omega^{*}_{|}}}\int_{[0,1]^d \setminus I_{\sim}}|\lambda([0,z))- \frac{1}{N}\sum_{i=1}^{N}\mathbf{1}_{[0,z)}(x_{i})|^{2}dzd\eta+\int_{I_{\sim}}\text{Var}( \frac{1}{N}\sum_{i=1}^{N}\mathbf{1}_{[0,z)\setminus [O',z)}(x_{i})) dz.
\end{aligned}
\end{equation}

For jittered sampling point set $P_{\Omega^{*}_{|}}=\{x_1,x_2,\ldots,x_N\}$,

\begin{equation}\label{J2}
    \begin{aligned}\mathbb{E}L_{2}^2(D_{N},P_{\Omega^{*}_{|}})&=\int_{P_{\Omega^{*}_{|}}}\int_{[0,1]^{d}}|\lambda([0,z))- \frac{1}{N}\sum_{i=1}^{N}\mathbf{1}_{[0,z)}(x_{i})|^{2}dzd\eta\\&=\int_{P_{\Omega^{*}_{|}}}\int_{[0,1]^d\setminus I_{\sim}}|\lambda([0,z))- \frac{1}{N}\sum_{i=1}^{N}\mathbf{1}_{[0,z)}(x_{i})|^{2}dzd\eta\\&+\int_{P_{\Omega^{*}_{|}}}\int_{I_{\sim}}|\lambda([0,z))- \frac{1}{N}\sum_{i=1}^{N}\mathbf{1}_{[0,z)}(x_{i})|^{2}dzd\eta.
    \end{aligned}
\end{equation}

Besides,
\begin{equation}\label{J3}
    \begin{aligned}
    &\int_{P_{\Omega^{*}_{|}}}\int_{I_{\sim}}|\lambda([0,z))- \frac{1}{N}\sum_{i=1}^{N}\mathbf{1}_{[0,z)}(x_{i})|^{2}dzd\eta\\&=\int_{I_{\sim}}\text{Var}( \frac{1}{N}\sum_{i=1}^{N}\mathbf{1}_{[0,z)\setminus [O',z)}(x_{i}))dz+\int_{I_{\sim}}\text{Var}( \frac{1}{N}\sum_{i=1}^{N}\mathbf{1}_{[O',z)}(x_{i}))dz.
\end{aligned}
\end{equation}

Easy to obtain

\begin{equation}\label{J4}
\int_{I_{\sim}}\text{Var}( \frac{1}{N}\sum_{i=1}^{N}\mathbf{1}_{[O',z)}(x_{i}))dz=4\cdot \frac{1}{2^d}\cdot \frac{1}{N^3}-5\cdot \frac{1}{3^d}\cdot \frac{1}{N^3}.
\end{equation}

Therefore, from \eqref{410},\eqref{J1},\eqref{J2},\eqref{J3} and \eqref{J4}, we have the desired result.

\section{Conclusion}\label{conclu}

We study expected $L_2-$discrepancy under two classes of  partitions,
and we give explicit formulas, these results improve the expected $L_2-$discrepancy formula on jittered sampling. It should be pointed out that the results presented here for the Class of partition models I and II need to be improved, they are all based on the strict condition $N=m^d$ of grid-based partition. The expected discrepancy of stratified sampling under general equal measure partition models will be investigated in future research and we hope to give some more applications in function approximation.


\begin{thebibliography}{ }

\bibitem{Dick2005}J. Dick and F. Pillichshammer, On the mean square weighted $L_2$ discrepancy of randomized digital $(t,m,s)$-nets over $\mathbb{Z}_2$, \emph{Acta Arith.}, 117(2005), 371-403.

\bibitem{Dick2006}L. L. Cristea, J. Dick and F. Pillichshammer, On the mean square weighted $L_2$ discrepancy of randomized digital nets in prime base, \emph{J. Complexity}, 22(2006), 605-629.

\bibitem{Dick2014}J. Dick and F. Pillichshammer, Discrepancy theory and quasi-Monte Carlo integration. In: W. W. L. Chen, A. Srivastav, G. Travaglini(eds.), Panoramy in Discrepancy Theory, Springer Verlag, Cham, 2014, 539-619.

\bibitem{Dick2020}J. Dick, A. Hinrichs and F. Pillichshammer, A note on the periodic $L_2$-discrepancy of Korobov's $p$-sets, \emph{Arch. Math.} (Basel), 115(2020), 67-78.

\bibitem{KP}M. Kiderlen and F. Pausinger, On a partition with a lower expected $L_2$-discrepancy than classical jittered sampling, \emph{J. Complexity}, 70(2022), https://doi.org/10.1016/j.jco.2021.101616.

\bibitem{KP2}M. Kiderlen and F. Pausinger, Discrepancy of stratified samples from partitions of the unit cube, \emph{Monatsh. Math.}, 195(2021), 267-306.

\bibitem{kirk2022expected}N. Kirk and F. Pausinger, On the expected $L_2$-discrepancy of jittered sampling, arXiv preprint arXiv:2208.08924, 2022.

\bibitem{PRS}F. Pausinger, M. Rachh, S. Steinerberger, Optimal jittered sampling for two points in the unit square, \emph{Statist.Probab. Lett.}, 132(2018), 55–61.

\end{thebibliography}
\end{document}